\documentclass[prb,twocolumn,showpacs,superscriptaddress,preprintnumbers,amssymb]{revtex4-2}
\usepackage{graphicx}
\usepackage{float}
\usepackage{color}
\usepackage{dcolumn}
\usepackage{bm}
\usepackage{physics}
\usepackage{comment}
\usepackage{svg}
\usepackage{placeins}
\usepackage{hyperref}
\usepackage{soul}

\newcommand{\beq}{\begin{equation}}
\newcommand{\eeq}{\end{equation}}
\newcommand{\beqn}{\begin{eqnarray}}
\newcommand{\eeqn}{\end{eqnarray}}

\newcommand{\cH}{ {\cal H} }

\newcommand{\cL}{ {\cal L} }

\newcommand{\cZ}{ {\cal Z} }

\newcommand{\hO}{\hat{O}}

\newcommand{\cx}[1]{{\color{black} #1}}

\newcommand{\nmj}[1]{{\color{black} #1}}

\newcommand{\abhi}[1]{{\color{black} #1}}

\newcommand{\sfigref}[2]{Fig.~\hyperref[#1]{\ref{#1}#2}}

\newcommand{\appref}[1]{Supplemental Material~\ref{#1}}

\begin{document}

\title{Probing Defects with Quantum Simulator Snapshots}

\author{Abhijat Sarma}

\affiliation{Department of Physics, University of California,
Santa Barbara, CA 93106, USA}

\author{Nayan Myerson-Jain}

\affiliation{Department of Physics, University of California,
Santa Barbara, CA 93106, USA}

\author{Yue Liu}

\author{Nandagopal Manoj}

\author{Jason Alicea}

\affiliation{Department of Physics and Institute for Quantum Information and Matter, California Institute of Technology, Pasadena, CA 91125, USA}

\author{Roger G. Melko}

\affiliation{Department of Physics and Astronomy, University of Waterloo, Ontario, N2L 3G1, Canada}

\affiliation{Perimeter Institute for Theoretical Physics, Waterloo, Ontario, N2L 2Y5, Canada}

\author{Cenke Xu}

\affiliation{Department of Physics, University of California,
Santa Barbara, CA 93106, USA}

\date{\today}

\begin{abstract}

Snapshots---i.e., projective measurements of local degrees of freedom---are the most standard data taken in experiments on quantum simulators, usually to probe local physics. In this work we propose a simple protocol to experimentally probe physics of defects with these snapshots. Our protocol relies only on snapshots from the bulk system, without introducing the defect explicitly; as such, the physics of different kinds of defects can be probed using the same dataset.
In particular, we demonstrate that with snapshots of local spin configurations of, for example, the $1d$ Rydberg atom realization of the quantum Ising criticality, we can (1) extract the ``defect entropy", and (2) access
the continuous line of fixed points of effective defect conformal field theory, which was recently discussed in the context of the ``weak-measurement altered criticality".

\end{abstract}

\maketitle


\section{Introduction}


Defects often hold the key to universal physics in both condensed matter and high energy theory that is not accessible \cx{through local probes alone}.
In various scenarios, local operators are incapable of fully characterizing the nature of the system, but extended objects like defects can provide a much sharper diagnosis. The most well-known example of such is the R\'{e}nyi entanglement entropy, which is mapped to a conical defect in the Euclidean spacetime path-integral~\cite{cardy1,cardy2}. Defects contain important universal information of the system. For example, the entanglement entropy contains information such as the central charge of a conformal field theory (CFT)~\cite{cardy1}, and quantum dimension of all the anyons of a topological order~\cite{kitaeventropy,wenentropy}. Another type of defect called the disorder operator encodes the information of the ``current central charge"~\cite{xudisorder,chengdisorder}, or the AC conductivity, which is a universal quantity at a $(2+1)d$ CFT~\cite{fisher1990}. It has also been shown recently that the physics of defects captures the universal physics of a class of problems in quantum information, \cx{such as measurement-induced phase transitions~\cite{QI_domainwall1,QI_domainwall2,QI_domainwall3}, and quantum systems under decoherence and weak-measurement~\cite{Ehud_2022,bao2023mixedstate,Lee_2023,sunyao,JianIsingWM,altered,Zou_2023,patil2024,Rajabpour_2016,Lin_2023,tim2025}.}
\cx{The study of defects in higher dimensional conformal field theories have also attracted broad interests from theoretical physics (see for example Refs.~\cite{defect1,defect2,defect3,zohardefect1,Casini_2019,Cuomo_2022,maxboundary,maxtoldin,Krishnan_2023,hedefect1,hedefect2,hedefect3,Wang_2022,wang2,linedefects3dising,Bill2016,Roy_2022,giombi2023,assaaddefect1}). And defects are tightly connected to the notion of ``generalized symmetries"~\cite{gensym1,gensym2}, one of the most dynamic directions of theoretical physics.}

Many highly entangled quantum many-body systems such as quantum spin liquids and exotic quantum critical points can be efficiently diagnosed with defects. However, so far defects have been used mostly as a numerical tool, rather than as an experimental probe. Direct experimental measurement of a nonlocal defect operator in condensed matter experiments is challenging. In this work, we propose that the defect operator can be quite conveniently and efficiently measured indirectly through snapshots of local degrees of freedom, which are the standard data routinely collected in experiments on quantum simulators. \cx{We note that recent studies have already explored physics of defects on various quantum simulators platforms~\cite{SLrydberg,stringgoogle,stringmaryland,stringgoogle}. Our work points out that there is a general simple protocol of probing defects across all quantum simulator platforms.} In particular, we propose the following experiments, which should be feasible with modern techniques. 

{\bf 1.} At a uniform $1d$ quantum critical state, such as the quantum Ising critical point of a Rydberg atom array, the snapshots of local spin configurations enables us to estimate the universal defect entropy, a notion proposed by Ref.~\onlinecite{ludwigentropy}, which encodes important information of the underlying conformal field theory. 

{\bf 2.} At the $1d$ quantum Ising critical point of Rydberg atom array, the snapshots of local spin configuration $\{ Z_j \}$ allows us to access the line of fixed points of effective CFT, which is related to the ``weak-measurement altered criticality", a direction that has attracted great interest in recent years. 


\section{Boundary/Defect Entropy}

We first discuss the boundary/defect entropy. Let us consider a quantum critical state, with a spatial boundary $A$, $B$ at $x = 0, L$. The partition function of the system with and without boundary (or periodic boundary) are labeled as $\cZ_{{\rm b}}$ and $\cZ$, which in the Euclidean spacetime path integral take the form: \beqn && \cZ = \int d \phi(x, \tau) \exp ( - \int d\tau dx  \ \cL ), \cr \cr && \cZ_{{\rm b}} = \int d \phi(x,\tau) \exp( - \int d\tau dx \ \cL - \int d\tau \ \delta \cH_{x = 0,L} ). \eeqn $\delta \cH_{x = 0,L}$ is the boundary Hamiltonian that is supposed to drive the boundary to a fixed point. The boundary entropy $\gamma$ introduced in Ref.~\onlinecite{ludwigentropy} is defined as \beqn \ln \langle \cZ_{{\rm b}}/\cZ \rangle = a \beta + \gamma_A + \gamma_B, \eeqn where $\beta = 1/T$ is the size of the system in the temporal direction. $\gamma_{A,B}$ are universal quantities independent of the microscopic details of the system, they are only controlled by the boundary fixed point. For example, if the CFT is a $(1+1)d$ 
critical Ising model, and the boundary Hamiltonian is a local Zeeman field that energetically favors polarizing the spin at $x = 0,L$, then the boundary corresponds to a standard Cardy state, and $\gamma_{A,B}$ is predicted to 
be $ \gamma_{A,B} = - (\ln 2) /2$. 
It was proven that $\gamma_{A,B}$ always decreases monotonically under renormalization group flow~\cite{entropyRG}, i.e. $\gamma_{UV} > \gamma_{IR}$.

\begin{center}
\begin{figure}[h!]
\includegraphics[width=0.45\textwidth]{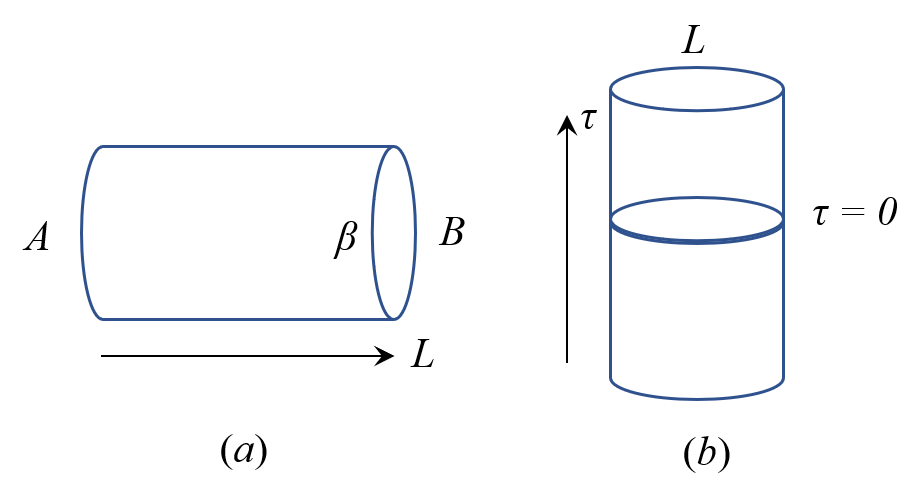}
\caption{$(a)$ The boundary entropy arising from the boundaries $A$ and $B$. $(b)$ The temporal defect after the space-time rotation. } \label{bd2}
\end{figure}
\end{center}


To facilitate experimental measurement, we map the boundaries to a ``temporal defect", in the Euclidean spacetime.  
Let us consider the following quantity: \beqn \frac{\cZ_{{\rm def}}}{\cZ} &=& \frac{\int d \phi(\tau, x) \exp( - \int d\tau dx  \ \cL - \int dx \ \delta \cH_{\tau = 0} )}{\int d \phi(\tau, x) \exp( - \int d\tau dx  \ \cL)} \cr\cr\cr &=& \left\langle{ \exp(- \int dx \ \delta \cH) }\right\rangle. \eeqn The last line of the equation above is simply the ground state expectation value of operator $\hat{O} = \exp(- \int dx \ \delta \cH )$, which is a spatially uniform operator. When $\delta H_{\tau = 0}$ is relevant, it is expected to effectively cut the system into two halves with two boundaries $\tau = 0^-$ and $0^+$. \nmj{For all the simple defects that we consider in this manuscript, this will be the case. Discussion of the expected IR behavior for many other simple defects can be found in Ref.~\cite{defectsgeneralizedpinning}}. We can define a defect entropy $\gamma$ as \beqn \ln \langle \hat{O} \rangle = a L + \gamma, \eeqn and $\gamma = \gamma_A + \gamma_B$. Note that $\ln \langle \hat{O} \rangle$ will always have a leading term that scales with the system size $L$. \cx{The space-time rotation maps the partition function of the boundary or a spatially localized defect, to the expectation value of a nonlocal operator in a {\it uniform} system; there is no need to introduce a defect explicitly. } 

Directly measuring a nonlocal operator is challenging in experiment, but we will demonstrate with examples that the expectation value $\langle \hat{O} \rangle$ can be evaluated through snapshots of local degrees of freedom. 
As an example, we first consider a nearest neighbor quantum Ising model, tuned to the critical point: \beqn H = \sum_j  - J \left( Z_{j} Z_{j+1} + X_j \right). \eeqn
And we consider the following form of operator $\hat{O}$:
\beqn \hat{O} = \exp( - \delta  \sum_j  Z_j ). \label{Zeemandefect} \eeqn This operator corresponds to turning on nonzero Zeeman field on the defect line $\tau = 0$. Because the scaling dimension of $Z_j$ is $1/8$, this defect is a relevant perturbation at the $(1+1)d$ Ising CFT, and will drive the line $\tau = 0$ to a defect fixed point with pinned $Z_j =+ 1$.

\begin{figure}[H]
    \centering
    \includegraphics[width=0.8\linewidth]{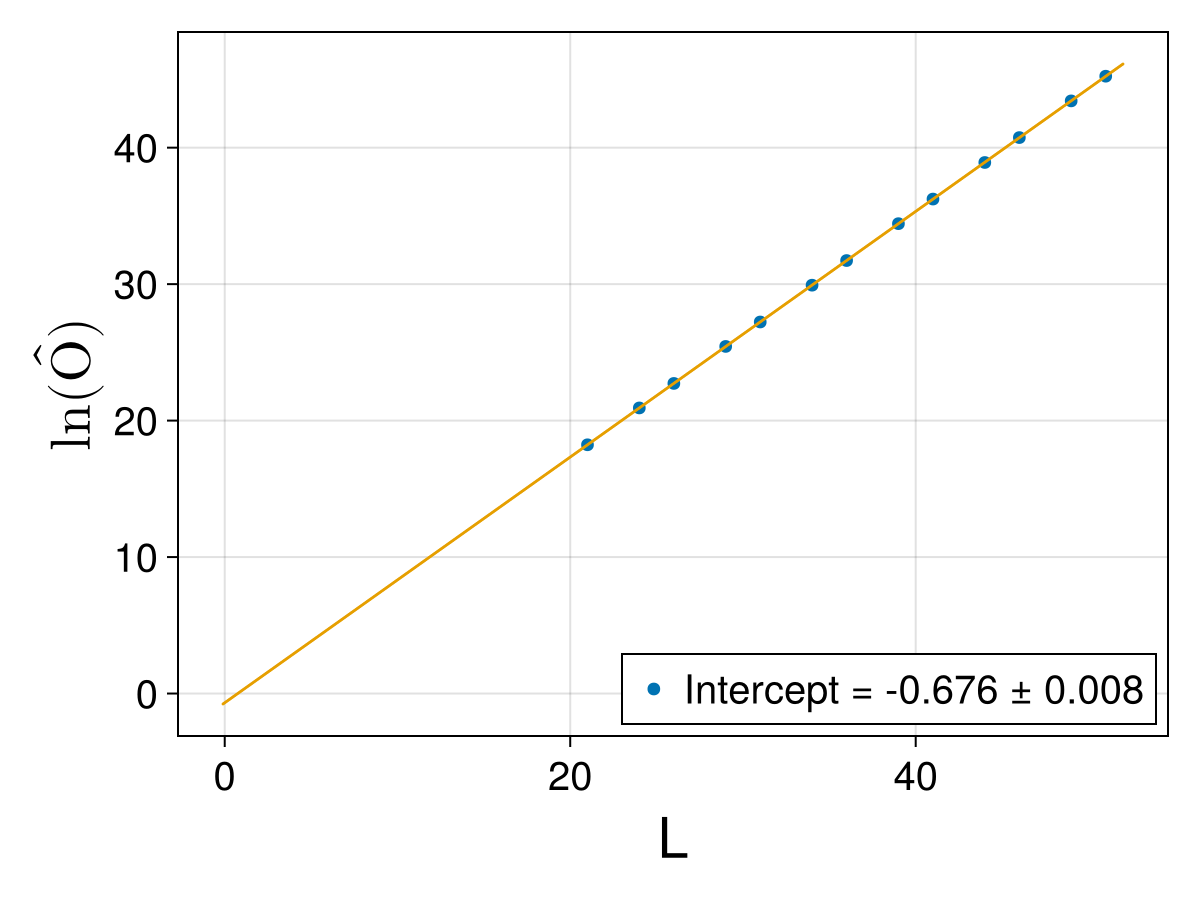} \caption{Critical Ising Defect Entropy at $\delta = 1$. The intercept gives $\gamma=-0.676 \pm 0.008$, while the theoretical value is $\gamma=-\ln{2} \approx -0.693$.}
    \label{fig:ising_gfac}
\end{figure}

\cx{We note here that the picture of mapping to temporal defect was previously used in the context of the interference of two Bose-Einstein condensates~\cite{interferedefect,interferedefect2}, where an intriguing relation was pointed out between the distribution of interference fringes and the defect partition function}. Here we propose that the expectation value of $\hat{O}$ can be efficiently evaluated through the ``snapshot Monte Carlo" that will be elaborated in the next section. Because $\hat{O}$ is positive-definite, Eq.~\eqref{snapshotO} should efficiently converge in the number of snapshots $M$. The defect entropy $\gamma$ is a universal value that only relies on the IR fixed point of the defect, so it does not depend on the bare value of $\delta$ in the thermodynamic limit. The theoretical value of $\gamma$ is $- \ln 2 \sim - 0.69$, which is in close agreement with our numerically extracted value in Fig.~\ref{fig:ising_gfac}. \cx{We note that the defect entropy of multiple copies of Ising chain also emerges when evaluating the Rényi entropy of decohered Ising CFT~\cite{Zou_2023,tim2025}. }


\section{Estimators from Snapshots}
Here we describe how snapshots can be used to estimate the expectation values of nonlocal operators. In particular, consider an operator $\hat{O}$ which is diagonal in some local basis denoted by $\ket{\{Z_j\}}$, $j=1,...,L$. Suppose we are given a dataset consisting of $M$ snapshots in this basis, $\{Z_j\}_m$, $m=1,...,M$, drawn from some probability distribution $p(\{Z_j\})$ determined by the Born rule. Such a dataset is the natural output of experimental runs on a quantum simulator. In the style of Monte Carlo, we can then construct an estimator for the underlying probability distribution based on the dataset: \beqn \tilde{p}(\{Z_j\}) = \frac{1}{M}\sum_{m=1}^M \delta_{\{Z_j\}, \{Z_j\}_m} \eeqn
\noindent This allows for us to estimate the expectation value of $\hat{O} = \sum_{\{Z_j\}} O(\{Z_j\})\ket{\{Z_j\}}\bra{\{Z_j\}}$: \beqn \label{snapshotO} \langle \hat{O} \rangle \approx \sum_{\{Z_j\}} \tilde{p}(\{Z\}) O(\{Z_j\}) = \frac{1}{M}\sum_{m=1}^M O(\{Z_j\}_m) \eeqn

In particular, this estimator relies only on local data, even if $\hat{O}$ is nonlocal, i.e. cannot be written as a sum of operators with bounded support. Further, the same dataset can be used to estimate the expectation value of various operators, as long as they are diagonal in the same basis. 
We make no assumptions about the source of the dataset, so long as the snapshots are sampled according to the Born rule; in particular, we use Stochastic Series Expansion (SSE), a worldline Monte Carlo technique \cite{Melko_SSE_2013, Merali_2024}, to generate the snapshots used for the data in this work, though the same procedure can be carried out using experimental data. We sample $M=10^6$ snapshots in each case. In \appref{App:C}, we show that the qualitative defect physics can be extracted even with $M\sim10^4$ snapshots, making it a realistic protocol for modern quantum simulators. We also discuss the theoretical sample complexity of our protocol in various limits in \appref{App:A}. 

One can use the snapshots protocol to measure defects coupling multiple copies of the system of interest as well. For example, consider a defect coupling two copies of the system, described by $\hat{O} = \sum_{\{Z_j^1,Z_j^2\}} O(\{Z_j^1, Z_j^2\})\ket{\{Z_j^1, Z_j^2\}}\bra{\{Z_j^1, Z_j^2\}}$, where $\{Z_j^1, Z_j^2\}$ denotes a local basis of the doubled system, and it is assumed that the two copies are uncoupled in the absence of the defect. First, we split the $M$ snapshots obtained from a single copy of our system into two sets of $M/2$ snapshots, with the configurations in each set labeled as $\{Z_j^1\}_m$, $\{Z_j^2\}_m$, respectively. Then $\langle \hat{O} \rangle$ can be estimated as $\langle \hat{O} \rangle \approx \frac{1}{M/2}\sum_{m=1}^{M/2} O(\{Z_j^1, Z_j^2\}_m)$. Thus, defects coupling multiple copies of the system can be measured using only snapshots from a {\it single copy}. This protocol could be used to measure the effect of a line defect coupling the spins of two critical Ising models, which describes the Renyi-2 physics of a critical Ising model with spin dephasing~\cite{Zou_2023}.

We note that it is possible to use snapshots to estimate off-diagonal operators as well in certain situations, in particular for so-called ``stoquastic" Hamiltonians, in which all of the amplitudes of the ground-state wavefunction can be taken to be real \cite{Torlai_2019, rydberggpt}. By training a generative model on the snapshots to learn the full distribution of measurement outcomes, one can extract the full ground-state wavefunction and use it to estimate off-diagonal observables. We do not pursue this avenue in this work.


\section{Rydberg Array}

\subsection{Defect Entropy}

The Rydberg array is a highly controllable experimental platform, well suited for quantum simulation applications. The system consists of a lattice of atoms whose electronic states can be driven into a highly excited ``Rydberg" state, with each atom forms a two-level system consisting of whether the Rydberg state is occupied or not. Atoms in the Rydberg state interact via $1/r^6$ dipole-dipole interactions which disfavor nearby atoms from both being in the Rydberg state. The Rydberg Hamiltonian is given by 

\begin{equation}
    H = \sum_{i<j} \frac{\Omega R_b^6}{|x_i - x_j|^6}+ \frac{\Omega}{2}\sum_j \sigma_j^x - \Delta \sum_j n_j
\end{equation}

\noindent where $n_j$ is 1 if the $j$-th atom is in the Rydberg state and zero otherwise, while $\sigma_j^x$ flips the value of $n_j$. Here, $R_b$, known as the blockade radius, is the characteristic length scale within which two atoms are strongly disfavored to both be in the Rydberg state. The parameters $\Delta$ and $\Omega$ are referred to as the detuning and Rabi frequency, respectively, and are akin to longitudinal and transverse fields in the usual Ising model. There are many ordered states in the phase diagram of the model depending on the dimensionality and the choice of parameters \cite{Bernien_2017, Merali_2024}. We are interested in tuning a transition between a $\mathbb{Z}_2$ charge density wave with occupancies $n_j = 1$ on every other site (an AF order of $n_j$), and a symmetric state where single-site translation is unbroken, which is expected to lie in the same universality class as the Ising transition. \cx{Scaling behaviors of the Ising criticality were recently observed in a Rydberg atom array~\cite{normanising}.} We fix the parameters $R_b$ and $\Omega$ in a region of the phase diagram which hosts the $\mathbb{Z}_2$ ordered state, treating $\Delta$ as the tuning parameter across the transition. \cx{In particular, we choose $a = 5.48\mu$m following the guidance of Ref.~\cite{Bloqade} and set $R_b = 1.920\cdot a$, where $a$ is the lattice constant. 
We consider both open and periodic boundary conditions, and find that the critical point is located at $\Delta_c/\Omega\approx 1.010, 1.015$ respectively. }

\begin{figure}[H]
    \centering
    \includegraphics[width=0.8\linewidth]{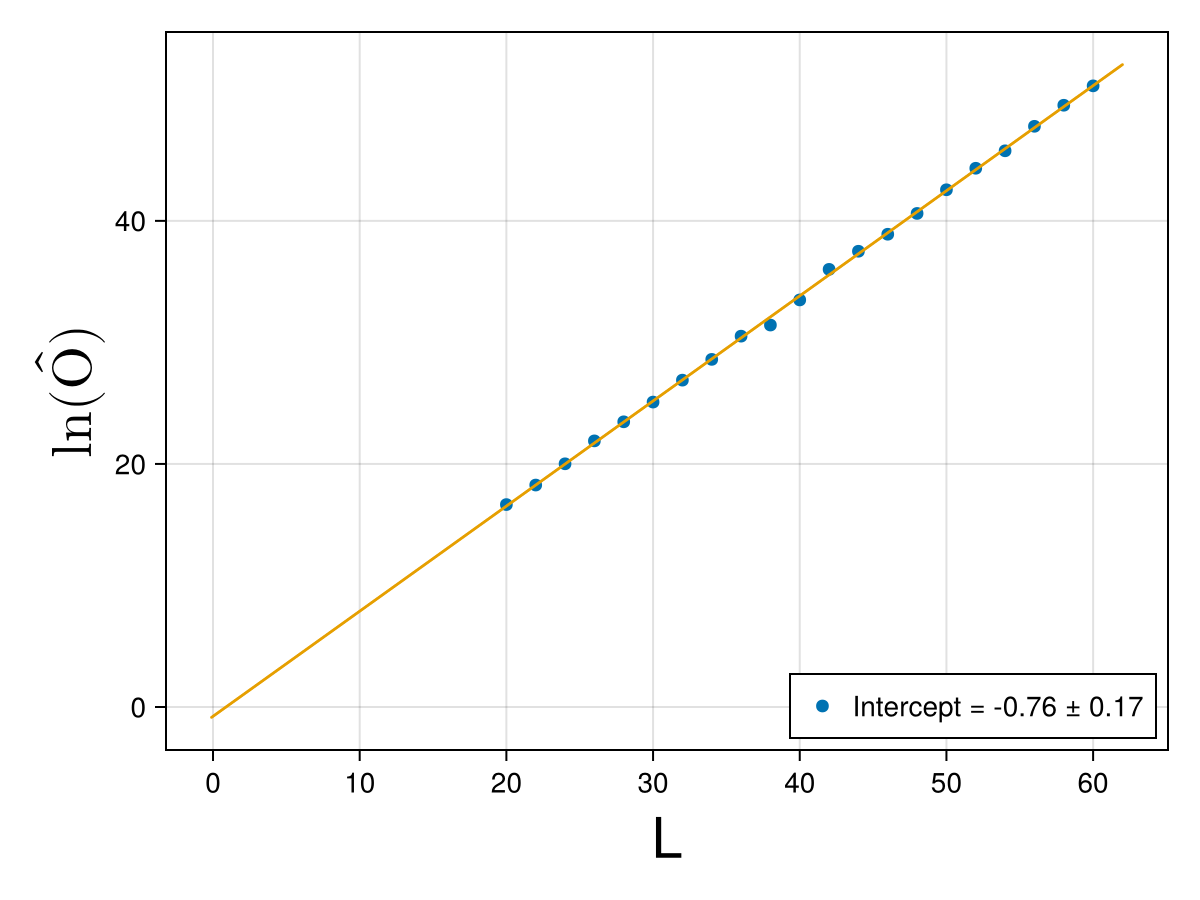} \caption{Critical Rydberg Defect Entropy at $\delta = 1$. The intercept gives $\gamma=-0.76 \pm 0.17$, while the theoretical value is $\gamma=-\ln{2} \approx -0.693$.}
    \label{fig:rydberg_gfac}
\end{figure}

\abhi{We can use the Rydberg array tuned to it's critical point as another verification of the defect entropy. 
We find that open boundary conditions pollutes the defect entropy at our system sizes, and so we use periodic boundary conditions to extract the defect entropy, which has been realized in Rydberg atom experiment~\cite{normanising}. As shown in Fig.~\ref{fig:rydberg_gfac}, we can still extract the defect entropy, though the result is less accurate compared with nearest neighbor Ising model, likely due to the long-range interactions in the Rydberg model. }

\subsection{Line of Fixed points}

In recent years the physics of quantum many-body systems under measurement and decoherence has attracted great interest. 
One example of such is quantum criticality under measurement and decoherence~\cite{JianIsingWM,altered,Zou_2023,sunyao}, and it was shown in recent studies that measurement can alter the critical behaviors, leading to unique novel critical  phenomena beyond any states of matter in equilibrium. Generally speaking, weak-measurement is mapped to a defect in the Euclidean spacetime path integral, and the physics under weak-measurement can be inferred from our understanding of conformal defects. For example, it was shown that equal-time bulk correlation of a $(2+1)d$ Wilson-Fisher quantum critical point is driven into the so-called ``extraordinary-log" phase~\cite{Lee_2023} that was discovered recently in the context of boundary CFT~\cite{maxboundary,maxtoldin,Krishnan_2023}.  

\begin{figure}
    \includegraphics[width=0.8\linewidth]{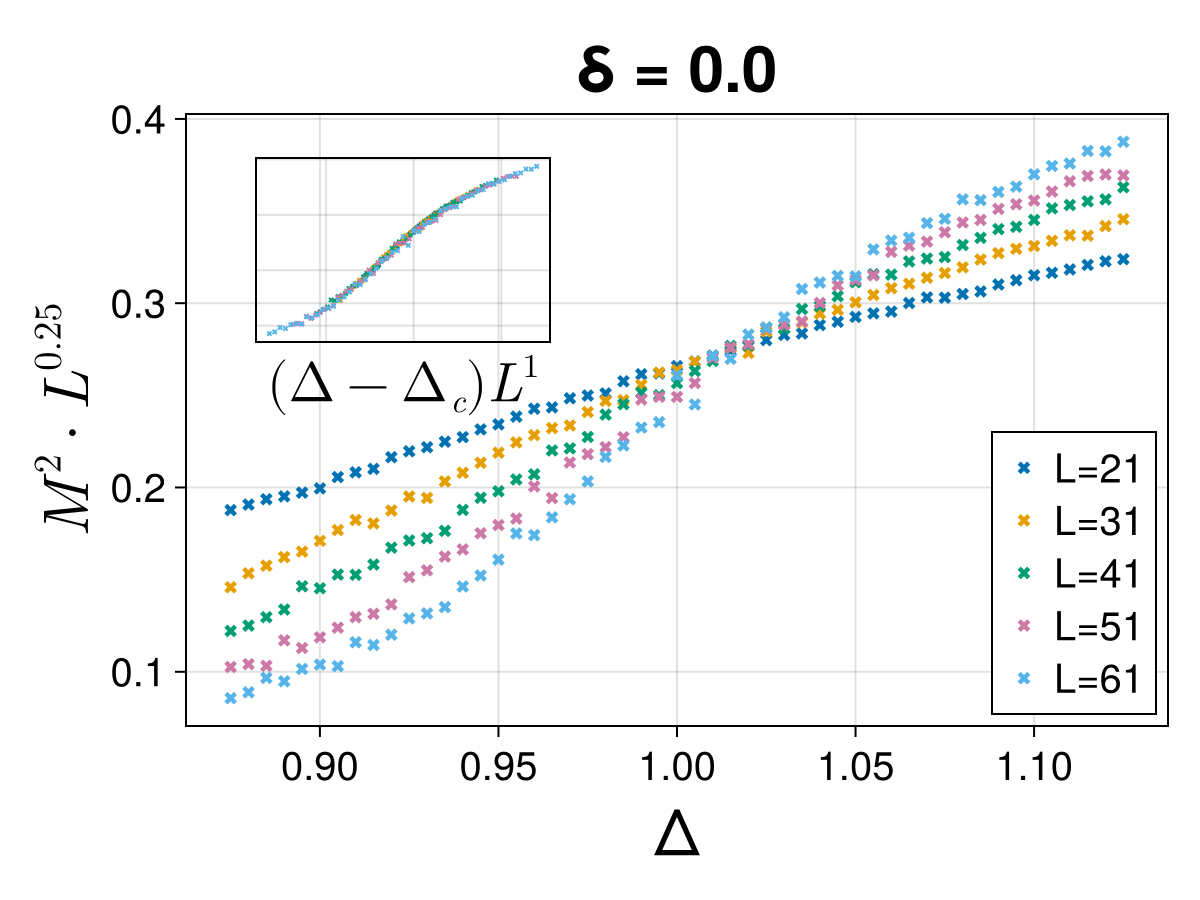}
    \includegraphics[width=0.8\linewidth]{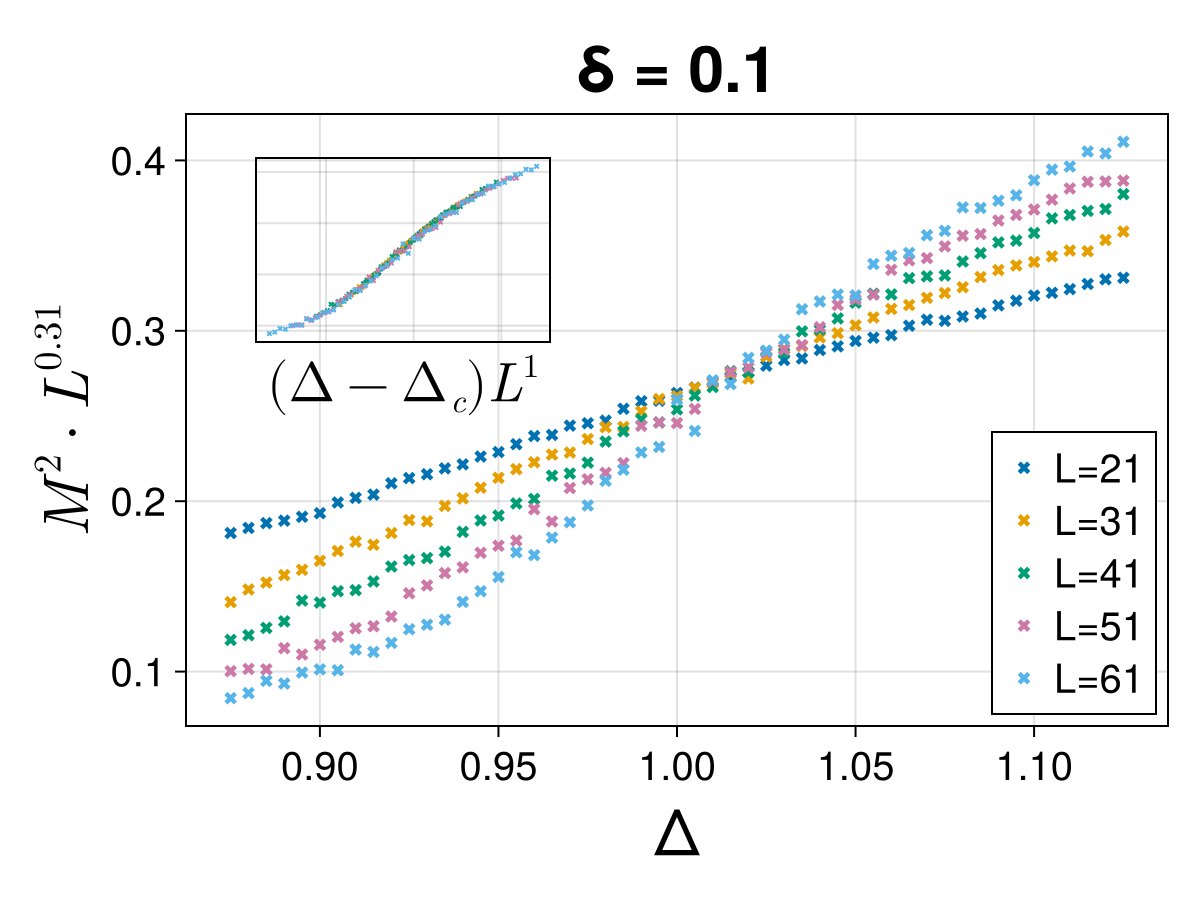}
    \caption{Line of Fixed Points (Continuously Varying Scaling Dimension). The quantity $M^2 L^{2D_d(\delta)}$ is scale invariant at the critical point, where $D_d(\delta)$ is the scaling dimension of the Ising spin field at the critical point along the defect line. For the bulk system given by $\delta=0$, we fix $D_d(0)=1/8$ and locate the critical point where the data for various system sizes crosses, giving $\Delta_c \approx 1.010$. For other choices of $\delta$, we choose $D_d(\delta)$ to minimize the variance of $M^2 L^{2D_d(\delta)}$ at $\Delta_c$. The numerically extracted values of $D_d(\delta)$ using this method agree closely with the theoretical prediction Eq.~\eqref{fp_formula}. The insets show the universal scaling collapse characteristic of criticality. More choices of $\delta$ are shown in Fig.~\ref{fig:fp_line_app} in SM.}
    \label{fig:fp_line}
\end{figure}

Within all the measurement-induced physics, here we focus on the ``line of fixed points" induced by weak-measurement on a $(1+1)d$ Ising CFT \cite{JianIsingWM,altered,Zou_2023,sunyao}. 
For instance, by weakly measuring the coupling between the spins, i.e., the nearest-neighbor two-body spin operator $Z_j Z_{j+1}$ 
 and post-selecting the outcome, the resulting physics is expected to be equivalent to the insertion of the temporal defect operator $\hat{O}$ 
\beqn
\hat{O} = \exp (- \delta \sum_{j} Z_{j} Z_{j+1} ) \label{Marginaldefect1}
\eeqn
In the continuum, the insertion of Eq.~\ref{Marginaldefect1}, amounts to the insertion of a temporal defect in the Ising CFT of the energy density $\delta \mathcal{H} \sim + \delta \varepsilon$. The energy density, $\varepsilon$, is a primary in the Ising CFT with dimension $\Delta_{\varepsilon} = 1$. Operators on the lattice have an expansion in terms of primaries (and descendants) of the CFT, for which the leading nontrivial primary for $Z_j Z_{j+1}$ is the energy density (see, e.g., Ref.~\onlinecite{IsingCFTLatticeExp}) $Z_j Z_{j+1} \sim 1 + \varepsilon + \dots$ As the energy density is exactly marginal, the scaling dimension of the spin operator $D_d(\delta)$ is a continuous function of the defect strength $\delta$ which controls the corresponding line of (defect) fixed-points \cite{MasakiDefect}. 

\cx{We emphasize again that in our protocol there is {\it no need} to actually postselect the outcomes, we simply need the snapshots of $\{ n_j \} $ of a critical Rydberg array, and the line of fixed points will emerge when we process the snapshots as follows}. The correlation functions we are interested in on the lattice to probe this defect physics takes the form
\beqn 
C^d(x, x', \delta) &=& \frac{ \langle Z_x Z_{x'} \exp(  - \delta \sum_{j = 1}^L Z_j Z_{j+1}) \rangle }{\langle \exp(  - \delta \sum_{j = 1}^L  Z_j Z_{j+1}) \rangle  }, \cr\cr
&\sim& \frac{1}{|x - x'|^{2D_d(\delta)}}.
\eeqn 
\abhi{In the above equation, 
$Z_j = (-1)^j (n_j - \langle n \rangle)$ \cite{Slagle_microscopic_rydberg}.} \abhi{Theory predicts that the the scaling dimension $D_d(\delta)$ is given by} \cite{BarievEnergyDefect, McCoyEnergyDefect} 
\begin{equation}
\label{fp_formula}
    D_d(\delta) = \frac{1}{8} + \frac{C \delta}{\pi} + \mathcal{O}(\delta^2) 
\end{equation}

\noindent where $C$ is proportional to the overlap of the lattice operator $Z_j Z_{j+1}$ with the $\epsilon$ field of the emergent Ising CFT, which is a nonuniversal quantity. The proportionality constant is fixed by $C=1$ for the critical nearest neighbor quantum Ising chain.

In Figs.~\ref{fig:fp_line} and \ref{fig:fp_line_app}, we plot the scaled quantity $M^2 L^{2D_d(\delta)}$ versus parameter $\Delta$ in the Rydberg atom model. \abhi{We find that the physics of the line of fixed points is more robust against different boundary conditions, and so we use open boundary conditions in our numerics, which are more easily realizable in experimental platforms.} The net squared magnetization $M^2$ of the system is defined as \beqn M^2 = \frac{1}{L^2}\frac{ \langle (\sum_j{Z_j})^2 \exp(  - \delta \sum_{j = 1}^L Z_j Z_{j+1}) \rangle }{\langle \exp( - \delta \sum_{j = 1}^L Z_j Z_{j+1}) \rangle  } \eeqn which is expected to have the same scaling with system size as $C^d(x, x')$ has with $|x-x'|$. \cx{We observe good data collapse consistent with scaling behaviors of the line of fixed points (Fig.~\ref{fig:fp_line},\ref{fig:fp_line_app}), meaning each $\delta$ is scaling invariant with its own scaling dimensions}, and the numerically extracted values of $D_d(\delta)$ are consistent with the theoretical prediction \abhi{(see Fig.~\ref{fig:scaling_dims_theory} in the SM)}. We note that since estimating both the defect correlation function and the net-squared magnetization involve using snapshots to calculate expectation values of non-local operators, the scaling complexity has the same features discussed in \appref{App:A} for the defect entropy. 


\section{Discussion}

\cx{We propose that various universal physics of conformal defects, such as the universal defect entropy, and the line of ``defect fixed points", can be probed in quantum simulators, through snapshots of local degrees of freedom of a uniform bulk CFT. This result is obtained through two observations: 

(1) Space-time rotation turns a spatially localized defect discussed in Ref.~\onlinecite{ludwigentropy} into a temporal defect, and the defect partition function is mapped to the expectation value of a nonlocal operator in a spatially uniform CFT.

(2) The expectation value of a nonlocal operator can be estimated through Monte Carlo on snapshots.

We tested our proposal with the transverse field quantum Ising model, as well as the Rydberg atom array. }

Our protocol can be generalized to higher dimensional systems as well, such as disorder operator which encodes important information of the CFT~\cite{hedefect1,xudisorder,chengdisorder,chengdisorder2}, and has been used as a numerical diagnosis for higher dimensional CFTs~\cite{chengdisorder,chengdisorder2,SMGdisorder}. We leave the discussion of probing conformal defects in higher dimensions to future exploration.

\section{Acknowledgements}

We thank Ejaaz Merali and Shaeer Moeed for developing and providing us with the SSE code for the Ising and Rydberg Array models used to generate the data in this work; it can be found at \url{https://github.com/PIQuIL/QMC_LTFIM}. We also thank Timothy Hsieh, David Mross, Sara Murciano, Stephen Naus, Pablo Sala, Xiangkai Sun, and Yijian Zou for helpful discussions.  C.X. and A.S. are supported by the Simons Foundation through the Simons Investigator program.  R.G.M acknowledges support from NSERC and the
Perimeter Institute. Research at Perimeter is supported in part by the Government of Canada through
the Department of Innovation, Science and Economic
Development Canada and by the Province of Ontario
through the Ministry of Economic Development, Job
Creation and Trade. The U.S. Department of Energy, Office of Science, National Quantum Information Science Research Centers, Quantum Science Center partially supported the sample complexity analysis of this work. 

\bibliography{Refs}

\FloatBarrier

\onecolumngrid

\appendix

\section{Sample Complexity}
\label{App:A}
Here we detail the sampling complexity of estimating the nonlocal observables considered in the main text using our snapshots protocol. For simplicity we focus throughout this appendix on the transverse-field Ising model at criticality; as discussed below, we expect similar conclusions to hold for the critical Rydberg chain as well. 

We first summarize our estimate of the sample complexity for parameter regimes explored in the main text.  

\begin{itemize}

    \item {\bf For the relevant pinning field defect $\delta H = \sum_j \delta Z_j $}

For the purpose of extracting the defect entropy we take $\delta = 1.0$. The required sample size in this case is $M \sim \exp( \alpha L ) / \epsilon^2 $, with $\epsilon$ denotes the desired accuracy, and $\alpha \approx 0.11$ in the transverse-fielding Ising model (and likely a comparable or smaller value for Rydberg arrays). Despite the exponential scaling with system size, the small value of $\alpha$ leads to modest required trials at experimentally relevant system sizes, e.g., for $L = 40$ and an error of $\epsilon = 0.05$ we have $M \sim 3\times 10^4$.  

\item {\bf For line of defect fixed points $\delta H = \sum_j \delta Z_{j} Z_{j + 1} $}

With system size $L \leq 50$, we take $\delta$ of order $0.1$. We see numerically that this is sufficiently large for these system sizes to access the exponents of the new fixed points. The scaling analysis in this Appendix shows that the relevant dimensionless parameter that controls the scaling is $\delta L^{1/2}$. 
$\delta L^{1/2} \ll 1$ implies polynomial sampling complexity with system size. Parameters $L = 50, \delta \sim 0.1$ give $\delta L^{1/2} = \mathcal{O}(1)$ where we are in the crossover regime between polynomial and exponential scaling, which is slow enough that we can estimate the non-local observable effectively with $10^6$ snapshots. For comparison, estimates using $10^4$ snapshots are shown in \appref{App:C}.

\end{itemize}

\subsection{Relevant line defect}

First consider the relevant magnetic line defect, corresponding to estimating 
\begin{equation}
    \langle{\hat{O}_{\mathrm{relevant}}}\rangle = \left\langle \exp\left(-\delta \sum_j Z_j\right)\right\rangle \qquad 
\end{equation}
with $\delta >0$ for concreteness. (In both the transverse-field Ising model and Rydberg array setup, the two signs of $\delta$ are symmetry-related.) 
To do so, we take $M$ snapshots of the critical wavefunction in the $Z$-basis---denoting the measurement outcomes $Z_{j;(m)} = \pm 1$ for the $m^{\rm th}$ run---and average via
\begin{equation}
    \langle{\hat{O}_{\mathrm{relevant}}}\rangle \approx \frac{1}{M} \sum_{m=1}^M \exp\left(-\delta \sum_j Z_{j;(m)}\right).
\end{equation}
We want to understand how well this average converges. In particular, how many measurement runs $M$ do we need to accurately estimate $\langle{\hat{O}_{\mathrm{relevant}}}\rangle$? 
Given the standard deviation of an operator $\sigma_{\hat{O}} = \sqrt{\langle \hat{O}^2\rangle-\langle\hat{O}\rangle^2}$, we know that after $M$ rounds of measurement the error in our estimate of $\langle{\hat{O}}\rangle$ goes as $\sigma_{\hat{O}} / \sqrt{M}$. If the relative error $\mathrm{RE} = \frac{\sigma_{\hat{O}}}{\langle{\hat{O}}\rangle} = \mathcal{O}(e^{a L})$ grows exponentially with system size $L$ for some constant $a>0$, then we require $M = \mathcal{O}(e^{2 a L})$, that is, exponentially many measurement rounds to estimate the expectation value to an $\mathcal{O}(1)$ relative error. In contrast, if the relative error scales only algebraically (or slower) with system size, then we only need polynomially many runs of the protocol to estimate the expectation value with similar precision. 

{\bf \emph{Small-$\delta$ limit.}}~In the `small-$\delta$' limit---which we quantify shortly---we can perform a cumulant expansion truncated at the leading nontrivial order to estimate the denominator in the relative error as $\langle \exp(-\delta \sum_j Z_j) \rangle \approx \exp(\frac{\delta^2}{2} \mathrm{Var}[\sum_j Z_j])$. The variance in the exponent evaluates to $ \mathrm{Var}[\sum_j Z_j] = \sum_{jk} \langle Z_j Z_k \rangle \sim L^{2-2\Delta_\sigma} = L^{7/4}$, where $\Delta_\sigma=\frac{1}{8}$ is the scaling dimension of the Ising spin, giving $\langle \exp(-\delta \sum_j Z_j) \rangle \sim e^{\kappa\delta^2L^{7/4}}$ for some constant $\kappa>0$. The truncated cumulant expansion exploited above is expected to hold provided $\delta L^{7/8} \ll 1$---which defines the small-$\delta$ regime.  (Notice that for any non-zero $\delta$, no matter how small, sufficiently large system sizes invariably escape this limit.)  Similarly evaluating the numerator of the relative error under the same assumptions yields $\mathrm{RE} \sim \sqrt{e^{2\kappa\delta^2 L^{7/4}} - 1} \sim \delta L^{7/8}$. 
Thus, the relative error grows only polynomially with $L$ in the small-$\delta$ regime. 

{\bf \emph{Large-$\delta$ limit.}}~In the limit where $\delta \gg 1$, the expectation value $\langle \exp(-\delta \sum_j Z_j) \rangle$ is instead dominated by the configuration where all Ising spins orient along the $-{\bf \hat{z}}$ direction. (For Rydberg arrays, where $Z_j = (-1)^j(n_j-\langle n\rangle)$, charge-density wave configurations dominate.) The probability of this maximally polarized configuration scales asymptotically like $p \sim e^{-\alpha L}$ with $\alpha \approx 0.11$ determined from fitting $\log(p)$ versus $L$ for critical transverse-field Ising chains with $L = [30, 40, 50, ..., 100]$. We thereby obtain $\langle \exp(-\delta \sum_j Z_j) \rangle \sim e^{(\delta - \alpha)L}$, in turn implying that the relative error grows exponentially in the large-$\delta$ limit: $\mathrm{RE} \sim e^{\frac{\alpha}{2} L}$, though importantly with relatively small $\alpha$.  Simulations from Ref.~\onlinecite{Naus} suggest that Rydberg arrays may exhibit even slower exponential scaling (see their Fig.~4).  

\begin{figure}
    \centering
    \includegraphics[width=\linewidth]{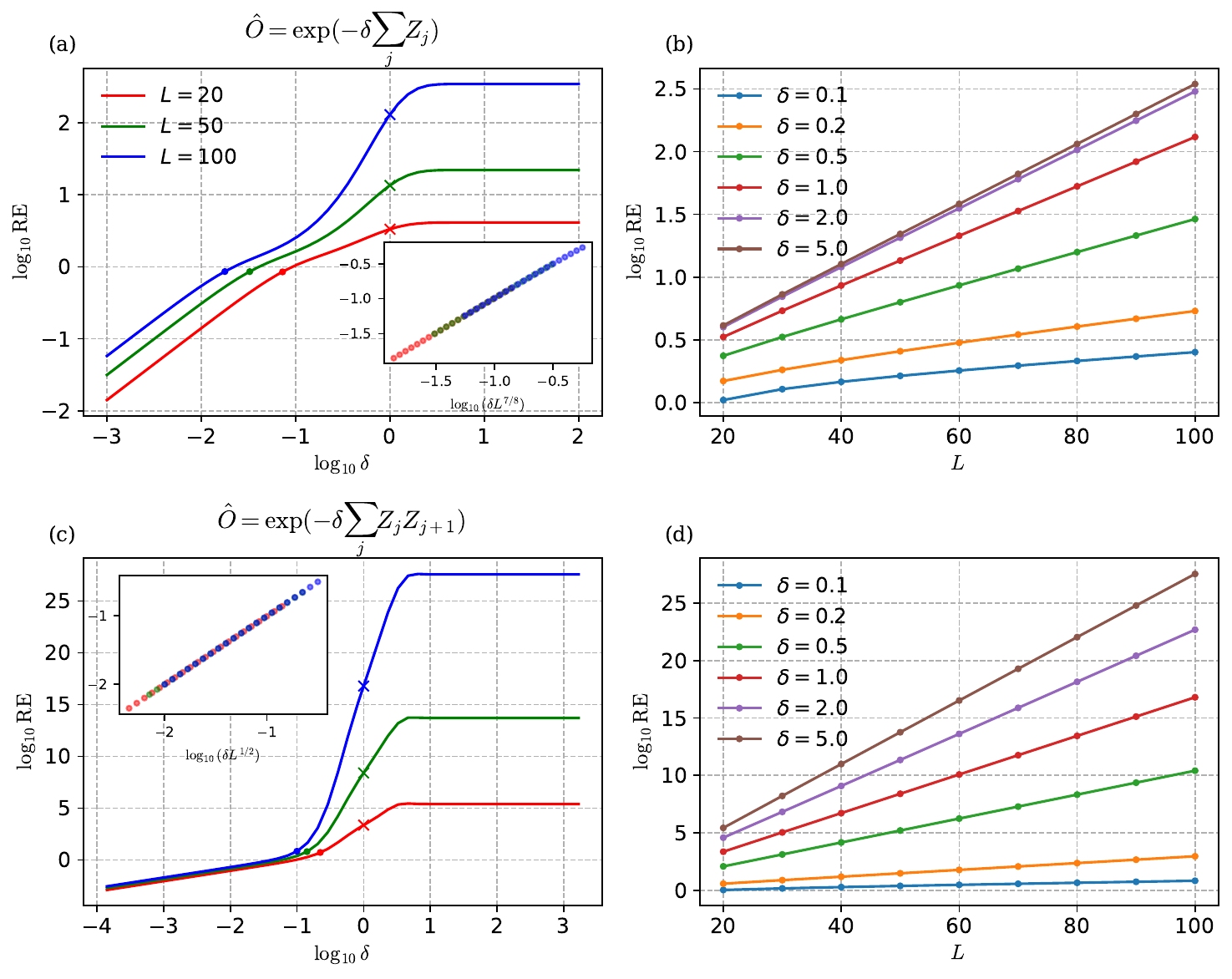}
    \caption{(a,c) Scaling of relative error (RE) versus defect strength $\delta$ for $Z_j$ and $Z_{j} Z_{j+1}$ defects. Dots in (a) and (c) mark the points with $\delta L^{7/8} = 1$ and $\delta L^{1/2} = 1$, respectively. The crosses mark the points with $\delta = 1$. Insets: $\log(\text{RE})$ versus $\delta L^{7/8}$ or $\delta L^{1/2}$ in the small-$\delta$ regime. (b,d) $\log(\text{RE})$ versus system size $L$ in the intermediate and large-$\delta$ regimes.}
    \label{fig:RE}
\end{figure}

{\bf \emph{Intermediate $\delta$.}}~We can also simulate the relative error numerically to capture the crossover between the two limits above.
Figure~\ref{fig:RE}(a) illustrates the $\delta$ and system-size dependence of the relative error 
obtained using density matrix renormalization group (DMRG) simulations of the transverse field Ising model.  Symbols on the curves roughly delineate endpoints of the small-$\delta$ and large-$\delta$ regimes (quantified by setting $\delta L^{7/8} = 1$ and $\delta = 1$, respectively).  Simulations recover the power-law system-size dependence for the former and exponential dependence for the latter; see the Fig.~\ref{fig:RE}(a) inset as well as Fig.~\ref{fig:RE}(b).  We also find numerically that the intermediate regime $L^{-7/8} \lesssim \delta \lesssim 1$, which we did not assess analytically, also exhibits exponential growth with $L$, albeit even slower than for the large-$\delta$ limit (see again Fig.~\ref{fig:RE}(b)).  

These results suggest that exponentially many experimental runs are required to accurately estimate $\langle{\hat{O}_{\mathrm{relevant}}}\rangle$ even if we fix $\delta \ll 1$ for sufficiently large $L$'s that place the system outside of the small-$\delta$ regime.

A relevant perturbation becomes increasingly visible in the IR, which operationally means that we must pick $\delta$ large enough so that the system can flow to the new defect CFT fixed point on the finite system size considered. In our problem we should therefore consider 
$\delta \sim O(1)$, where the sample complexity suffers from exponential growth based on the estimates above. However, this growth is still slow enough for our purposes (recall the small value $\alpha \approx 0.11$ for the large-$\delta$ limit).  In Figure \ref{fig:ising_gfac}---where we used $\delta = 1$---we fit transverse-field Ising chain data over a range of system sizes $\Delta L = 30$, meaning the error of the datapoint on the largest system size should only be bigger than the error of the datapoint on the smallest system size by a factor of roughly $e^{0.055\cdot30}=5.2$. In our numerics we find that this ratio is $4.5$, in close agreement with the scaling prediction. 

\subsection{Marginal line defect}

The situation for an exactly marginal perturbation differs qualitatively. Let us repeat a similar error analysis for $\langle \hat{O}_{\mathrm{marginal}}\rangle = \langle\exp(-\delta \sum_j Z_j Z_{j+1})\rangle$.  The two signs of $\delta$ are no longer symmetry-related; we choose $\delta>0$ but discuss differences with $\delta < 0$ below.  

{\bf \emph{Scaling analysis at $\delta > 0$.}}~For `small' $\delta$ we again perform a truncated cumulant expansion to obtain $\mathrm{Var}[\sum_j Z_j Z_{j+1}] = \sum_{jk} \langle Z_j Z_{j+1} Z_k Z_{k+1}\rangle_c \sim \sum_{j\neq k} \frac{1}{|j-k|^2} \sim L$, where we used the scaling dimension $\Delta_{\epsilon}=1$ of the energy field. The relative error follows as $\mathrm{RE} \sim \delta L^{1/2}$, indicating polynomial complexity when $\delta L^{1/2} \ll 1$ (which now defines the small-$\delta$ limit). \cx{Though for the purpose of observing the line of defects, it suffices to take $|\delta |$ of order 0.1, for completeness we also discuss sample complexity for $\delta \gg 1$.} The limit $\delta \gg 1$ still exhibits exponential growth via the same reasoning as above, i.e., $\mathrm{RE} \sim e^{\frac{\alpha'}{2} L}$.  
The prefactor $\alpha'$, however, differs from $\alpha$ introduced in the previous subsection since $\mathcal{O}_{\rm marginal}$ amplifies low-probability configurations at $\delta>0$ (antiferromagnetic states for the Ising model, or uniform states for Rydberg).  In particular, we will see below that $\alpha'$ significantly exceeds $\alpha$.  


For the marginal perturbation case, the non-local expectation value in the transverse field Ising model setting can be calculated exactly at any $\delta$ using fermionic Gaussian states~\cite{Surace_Tagliacozzo_2022,Fagotti_Calabrese_2010}. Figure~\ref{fig:RE}(c) illustrates the resulting relative error.  Once again our simulations recover power-law scaling for the small-$\delta$ regime and exponential scaling otherwise; see the Fig.~\ref{fig:RE}(c) inset and Fig.~\ref{fig:RE}(d).  Notice the much larger relative error at $\delta\gtrsim 1$ in Fig.~\ref{fig:RE}(c) compared to (a).  This distinction reflects the low-probability configurations amplified by $\hat{O}_{\rm marginal}$, in turn leading to significantly faster exponential growth with system size.  


For a marginal perturbation, there is no ``flow" necessitating $\delta$ to be large to see a noticeable effect on finite-size systems. Because marginal perturbations are equally important in the UV and IR, 
as long as the system size is large enough to see the bulk scaling dimension $\Delta_{\sigma} = \frac{1}{8}$, the modified scaling dimension $D_d(\delta)$ predicted by the line of fixed points should be visible. Therefore we are free to operate in the regime where $\delta \lesssim {\rm const} \times L^{-1/2} $, in which we can avoid the exponential sampling complexity and still measure the physics of the line of fixed points. Of course $\delta$ must still be large enough so that the modified scaling $1/|x-x'|^{2 D_d(\delta)}$ can be discriminated from $1/|x-x'|^{2 D_d(0)}$ on accessible length scales, which is possible if $L^{-2 (D_d(\delta) - D_d(0))}$ is far from 1, setting a lower bound $\delta \gtrsim {\rm const}'/(\ln L).$ Whether both of these inequalities can be simultaneously satisfied depends on the microscopic system under consideration, as the factors ${\rm const}$ and ${\rm const}'$ depend on microscopic details.  In our numerics on the Rydberg array we choose $|\delta| \approx 0.1$ and $L \approx 50$, which is close to the regime where polynomial scaling is expected, and we can clearly discriminate the different scaling dimensions on these system sizes, showing that our method can successfully measure the line of fixed points without exponentially many samples.

{\bf \emph{Differences at $\delta <0$.}}~At $\delta<0$ the same scaling behavior captured above continues to hold, i.e., power-law scaling with system size when $\delta L^{1/2} \ll 1$ and exponential scaling otherwise.  Nontrivial differences do appear, however, given that $\hat{O}_{\rm marginal}$ amplifies the highest-probability configurations at $\delta <0$ (all-up or all-down states for the transverse field Ising model, or charge-density-wave states for Rydberg arrays).  Consequences become particularly stark at $|\delta| \gg 1$.  The prefactor $\alpha'$ governing exponential scaling in this regime matches $\alpha$ from the previous subsection, i.e., the growth with system is much slower compared to $\delta > 0$. 
In turn, the relative error at $|\delta| \gtrsim 1$ (not shown) becomes greatly suppressed relative to Fig.~\ref{fig:RE}(c), yielding non-monotonic relative error as a function of $\delta$ that peaks when $|\delta| = \mathcal{O}(1)$.


\section{Extra data}
\label{App:B}
\begin{figure}[h]
    \centering
    \includegraphics[width=0.32\linewidth]{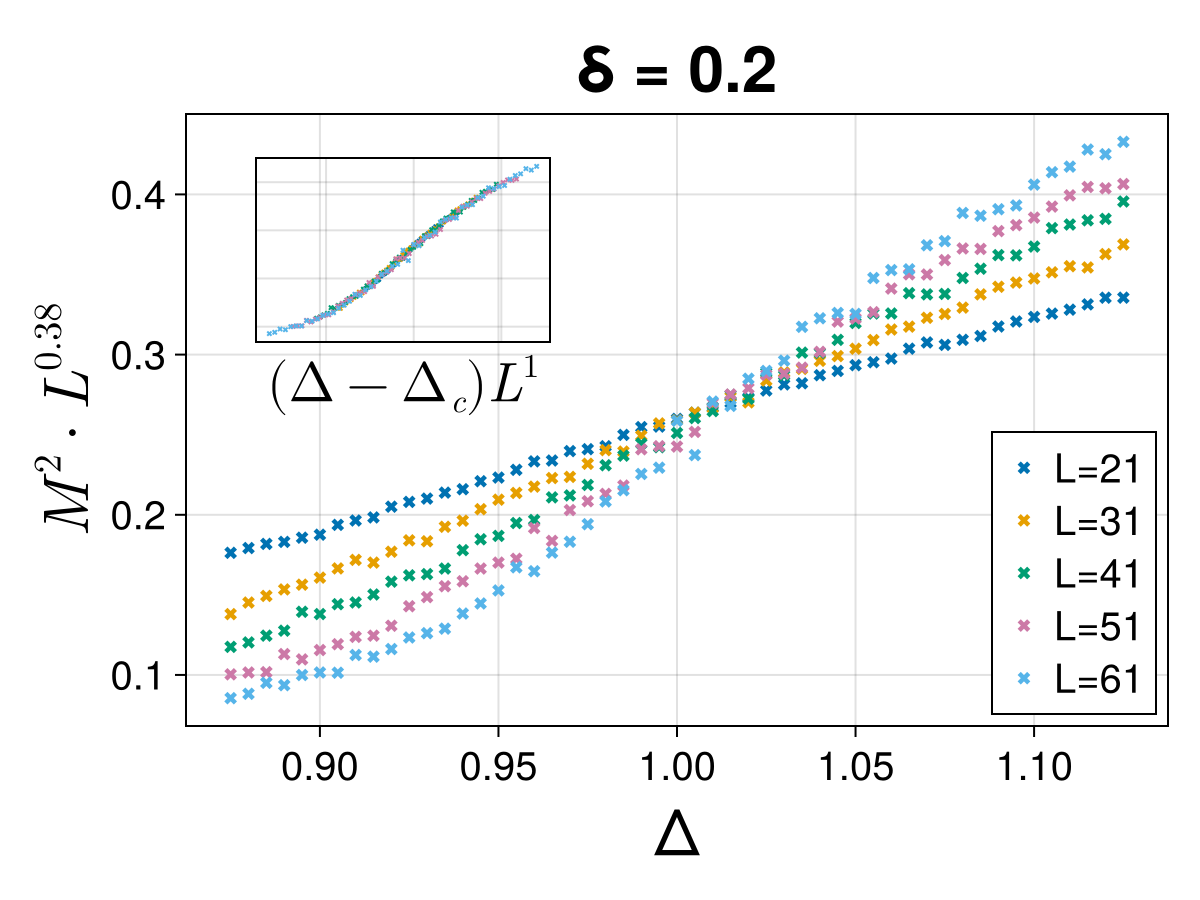}
    \includegraphics[width=0.32\linewidth]{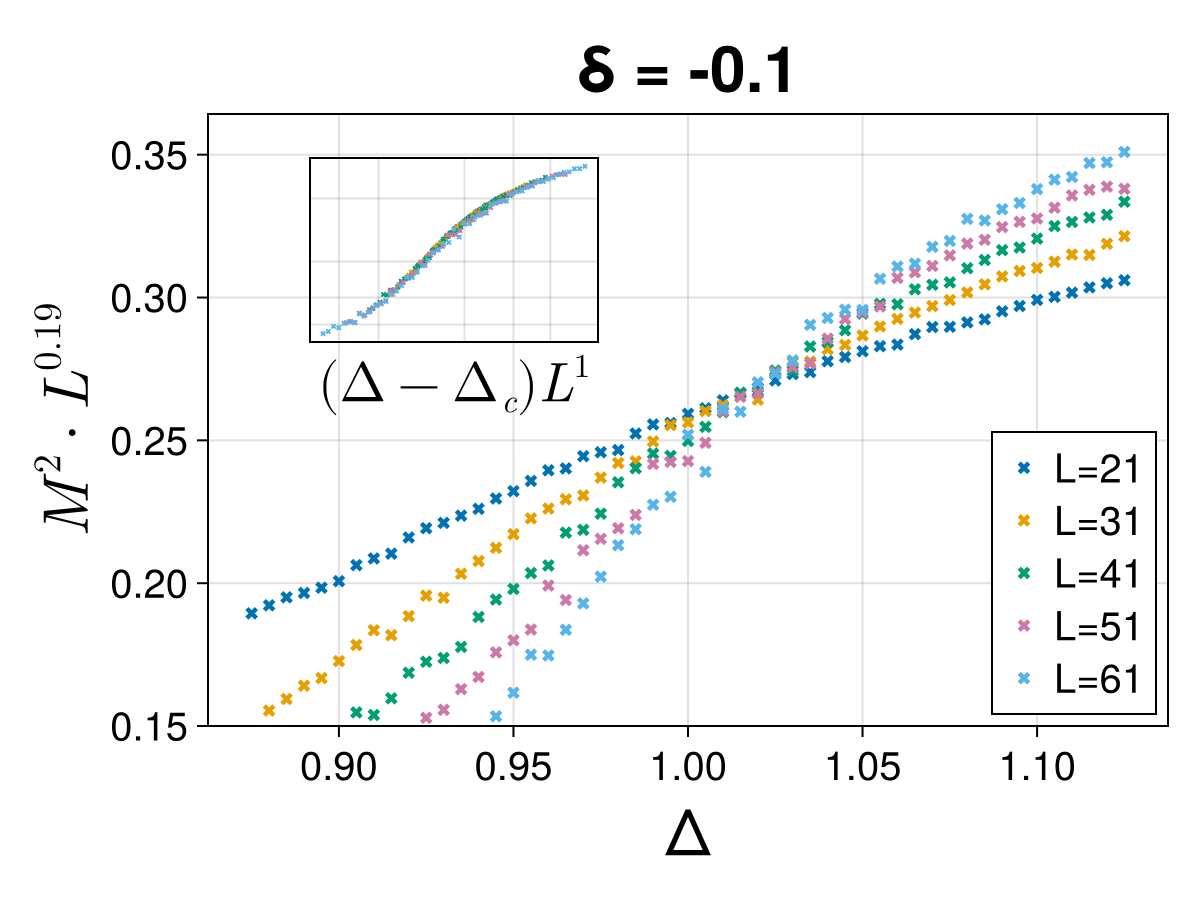}
    \includegraphics[width=0.32\linewidth]{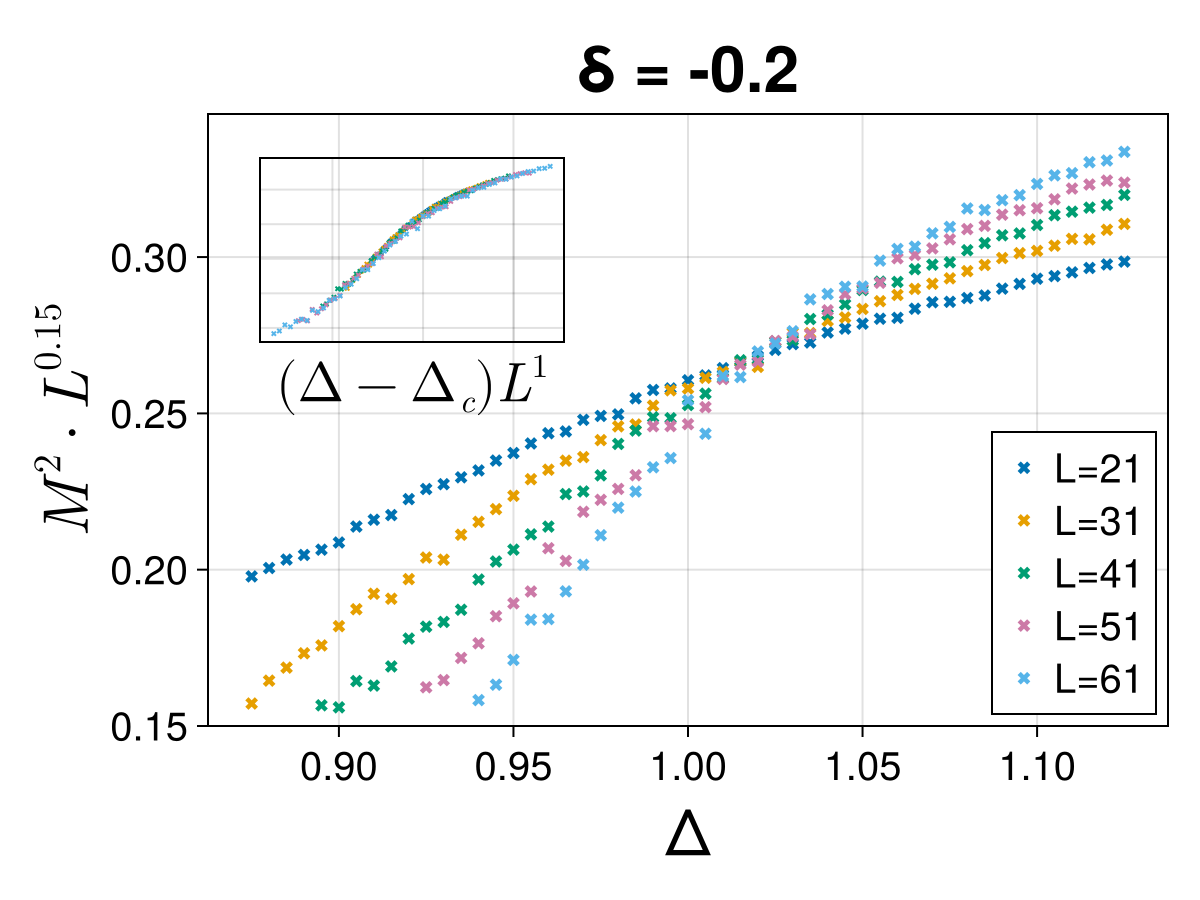}
    \caption{Line of Fixed Points (Cont.).}
    \label{fig:fp_line_app}
\end{figure}

\begin{figure}[h]
    \centering
    \includegraphics[width=0.4\linewidth]{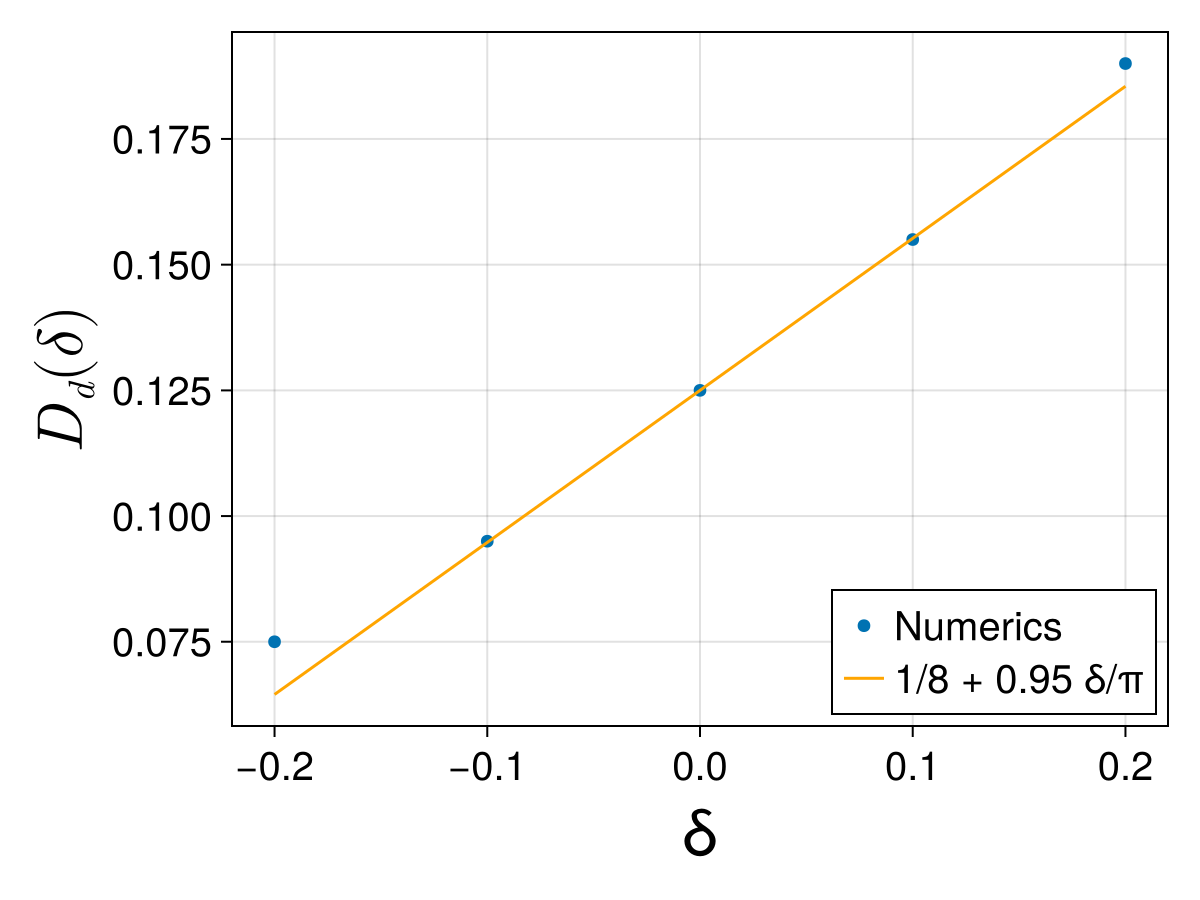}
    \caption{Line of Fixed Points Scaling Dimensions, Numerics vs Theory. We find good agreement for $C=0.95$ in the small $\delta$ regime where the theoretical formula is valid.}
    \label{fig:scaling_dims_theory}
\end{figure}

\begin{figure}[h]
    \centering
    \includegraphics[width=0.4\linewidth]{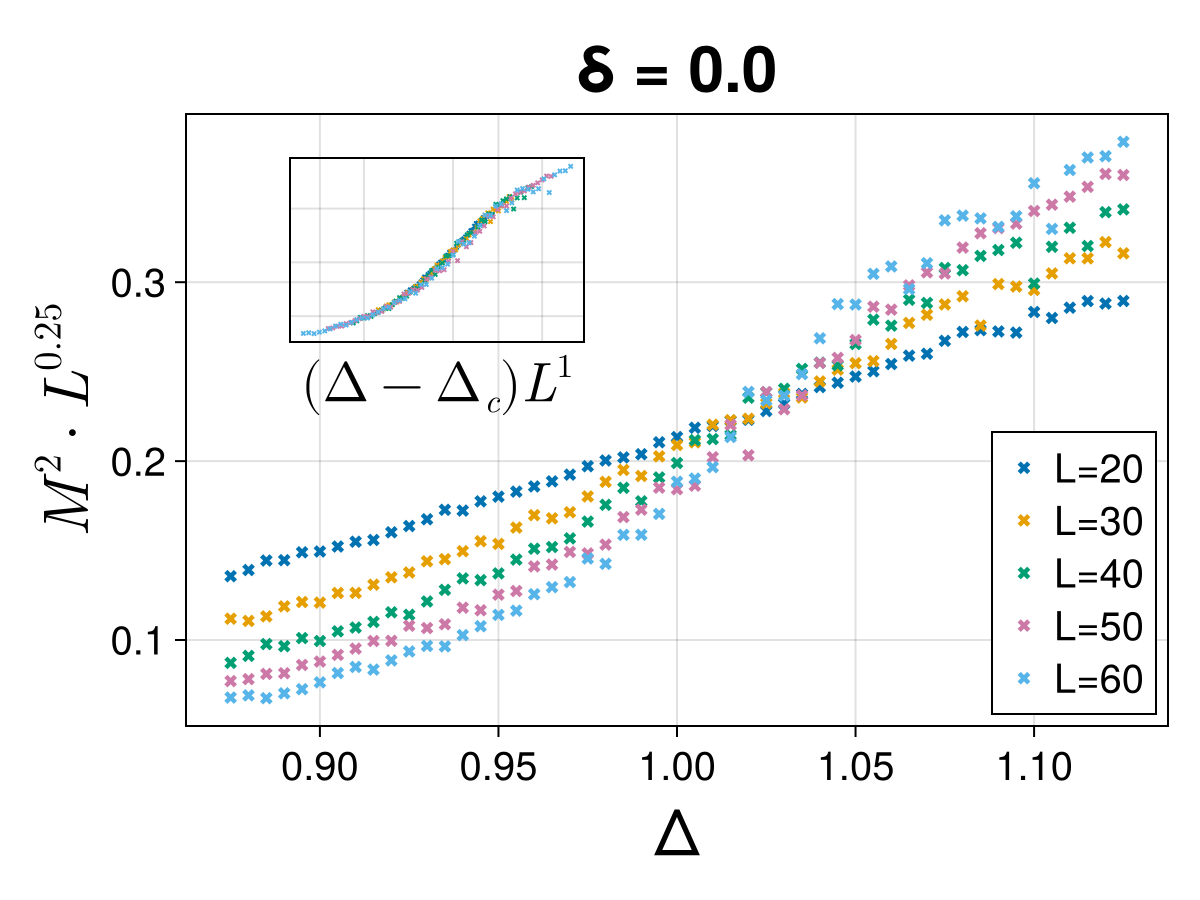}
    \caption{Rydberg Crossing Point Analysis, Periodic Boundary Conditions. The crossing point analysis is used to determine the location of the critical point as $\Delta_c \approx 1.015$.} 
    \label{fig:scaling_dims}
\end{figure}

In the main text we have discussed that, the operator $\hO$ in Eq.~\ref{Marginaldefect1} drives the system into a line of defect fixed points parameterized by $\delta$. Here we present more data with different choices of $\delta$ (Fig.~\ref{fig:fp_line_app}). For all $\delta$ the data $M^2 L^{2 D_{d}(\delta)} $ collapse with certain scaling dimension $D_{d}(\delta)$. And $D_{d}(\delta)$ increases monotonically with $\delta$, which is consistent with the physical picture that increasing $\delta$ weakens the antiferromagnetic Ising order. The key is that, for any strength of $\delta$, there is a different scaling dimension $D_{d}(\delta)$, featuring the line of defect fixed points. Theory predicts the value of $D_{d}(\delta)$ in Eq.~\ref{fp_formula} with $C=1$ for the quantum Ising chain. The extracted scaling dimensions from our numerics on the Rydberg array agrees well with the theory prediction with $C=0.95$. This suggests that the overlap of $Z_j Z_{j+1}$ with the $\epsilon$ field is quite similar in the two lattice models.

We also present more data for the Rydberg atom chain with periodic boundary, which is a geometry used in Ref.~\onlinecite{normanising} for the observation of Ising criticality. Based on the data the critical point for the system is determined at $\Delta_c \approx 1.015$.
 
\section{Limited Snapshot Data}
\label{app:limited}
\label{App:C}

\begin{figure}[h]
    \centering
    \includegraphics[width=0.4\linewidth]{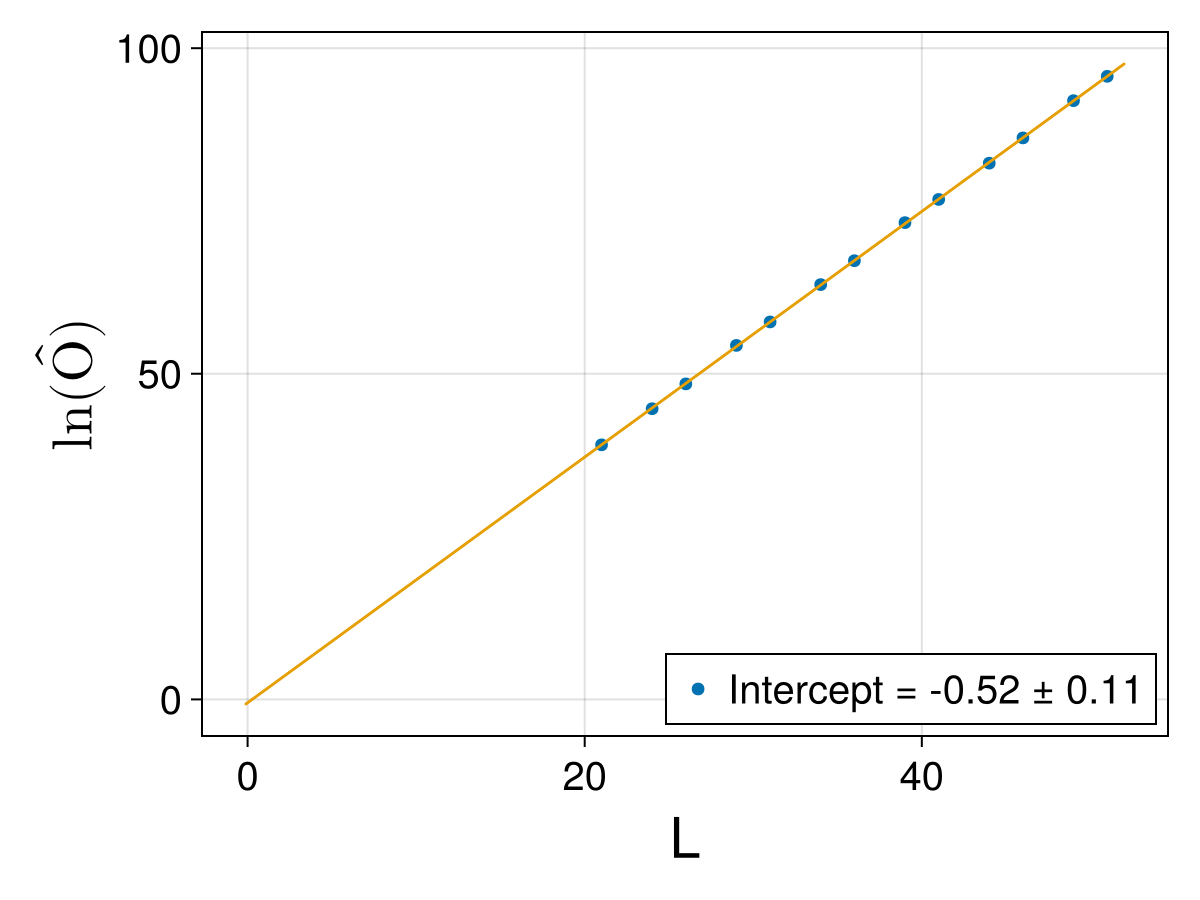}
    \caption{Critical Ising Defect Entropy, extracted with $M=10^4$ snapshots. The error from the theoretical value is $25\%$.}
    \label{fig:ising_gfac_limited}
\end{figure}

\begin{figure}[h]
    \centering
    \includegraphics[width=0.4\linewidth]{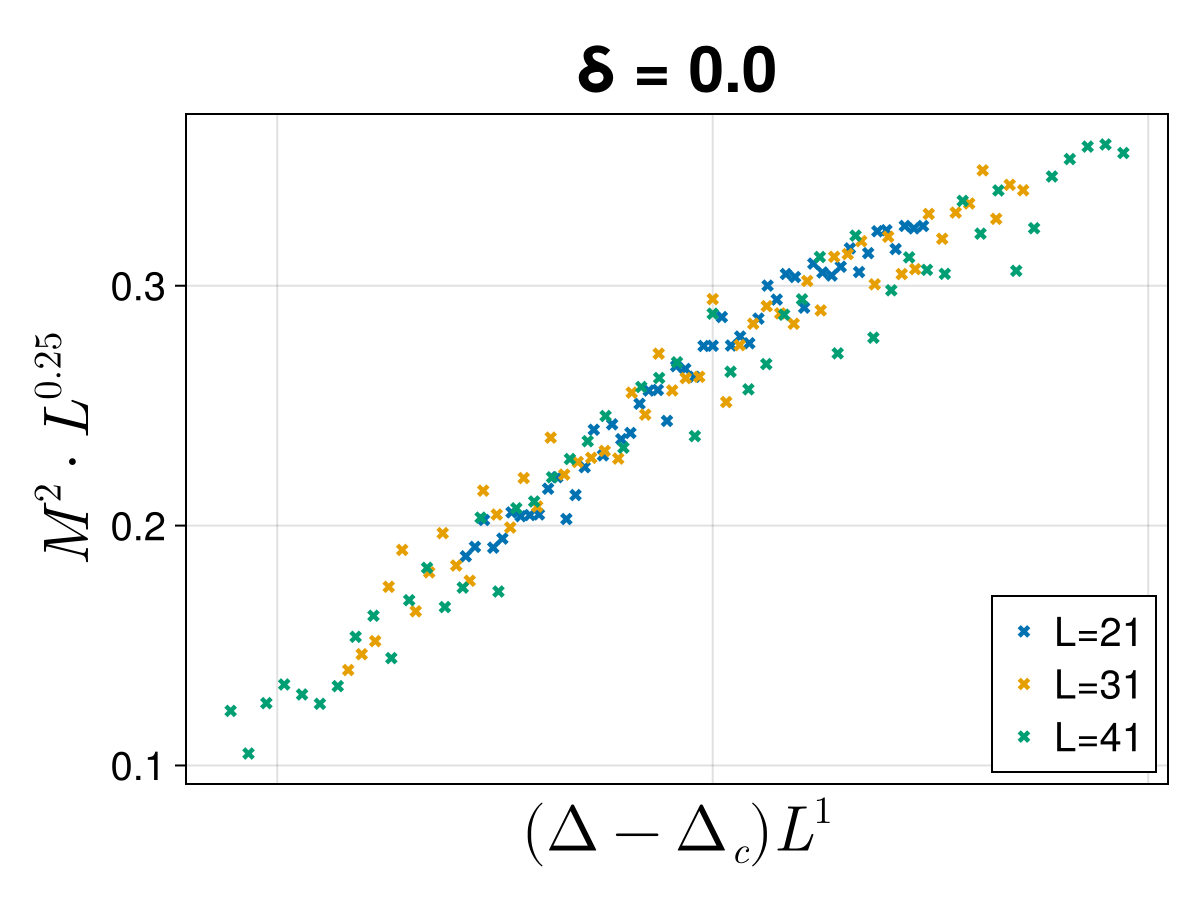}
    \includegraphics[width=0.4\linewidth]{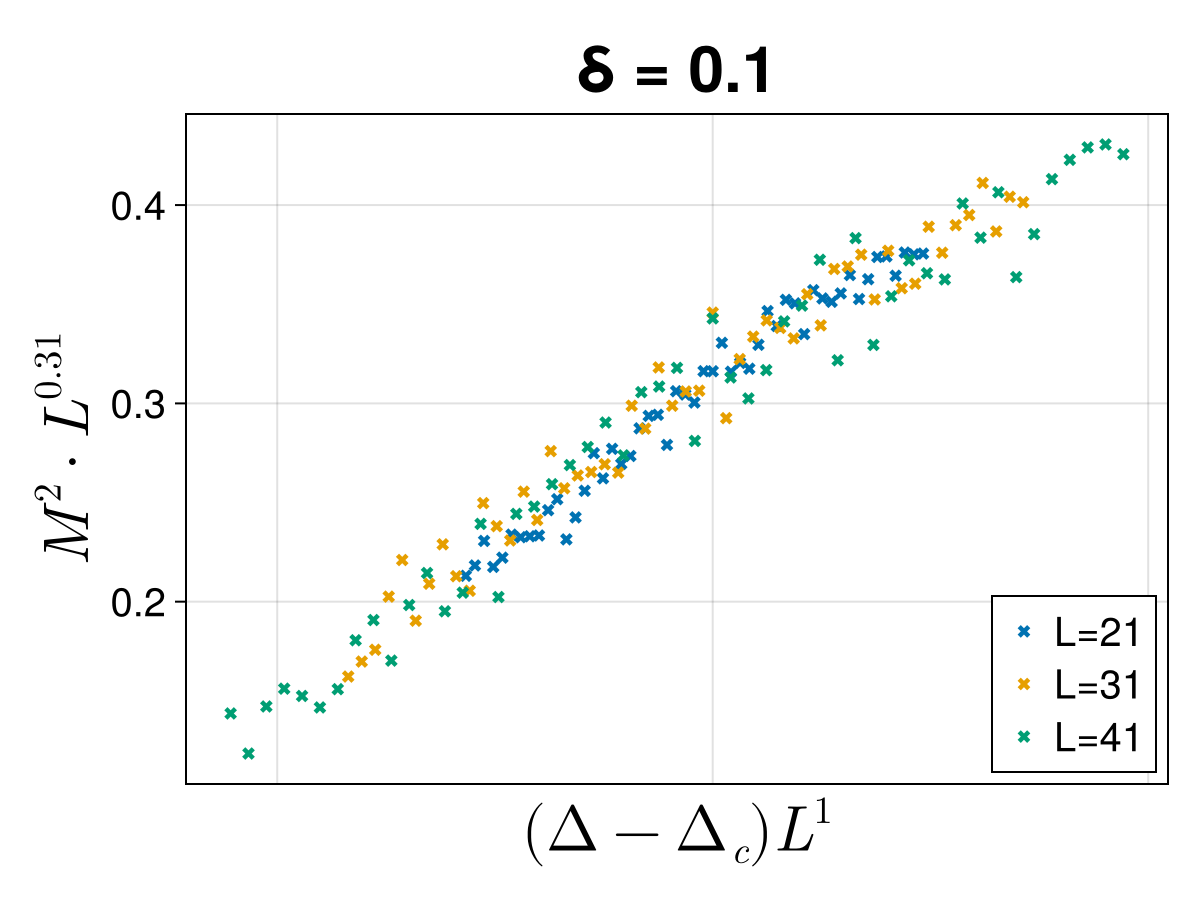}
    \caption{Line of Fixed Points, extracted with $M=2\cdot 10^4$ snapshots. Crossing point analysis is difficult with fewer snapshots, but we can still observe clear scaling collapse, allowing us to extract the fixed point and scaling dimensions.}
    \label{fig:fp_line_limited}
\end{figure}

In experimentally relevant settings, the number of snapshots that one can generate is often limited by resource constraints. As such, it is important to determine how reliably we can extract the quantities of interest using limited snapshot data. Here, we show that the qualitative aspects of the physics explored in this work can be observed even with $M \sim 10^4$ snapshots. 

As one can see, with 10K samples, the extracted defect entropy of the critical quantum Ising chain with temporal defect operator Eq.~\ref{Zeemandefect} and $\delta=0.1$ is off by about 25\% compared with the theoretical value. 

For the Rydberg atom model with an inserted marginal temporal defect Eq.~\ref{Marginaldefect1}, the exact critical point is difficult to determine with the order of $10^4$ samples. But the data collapse of $M^2 L^{2 D_{d}(\delta)}$ can still be seen.

\end{document}